\documentclass[runningheads]{llncs}
\usepackage[T1]{fontenc}
\usepackage{graphicx}
\usepackage{hyperref}
\usepackage{color}

\usepackage{newclude}

\usepackage{listings}
\usepackage{algorithm}

\newcommand{\lingo}[1]{\texttt{#1}}
\newcommand{\jargon}[1]{\lingo{#1}}

\usepackage{xspace}
\newcommand{\codename}[1]{\textsc{#1}}
\newcommand{\swift}{\codename{Swift}\xspace}
\newcommand{\swiftgpupacksim}{\codename{Swiftgpupacksim}\xspace}

\newcommand{\sphenix}{\texttt{Sphenix}\xspace}
\newcommand{\eagle}{\texttt{Eagle25}\xspace}
\newcommand{\gresho}{\texttt{Gresho256}\xspace}

\newcommand{\dine}{\texttt{Intel+A30}\xspace}
\newcommand{\grace}{\texttt{Grace~Hopper}\xspace}

\newcommand{\packing}{\jargon{packing}\xspace}
\newcommand{\launch}{\jargon{launch}\xspace}
\newcommand{\unpacking}{\jargon{unpacking}\xspace}
\newcommand{\density}{\jargon{density}\xspace}
\newcommand{\gradient}{\jargon{gradient}\xspace}
\newcommand{\force}{\jargon{force}\xspace}

\newcommand{\partstruct}{\jargon{part-struct}\xspace}
\newcommand{\globalvar}{\jargon{global-var}\xspace}
\newcommand{\explicitvar}{\jargon{explicit-var}\xspace}

\newcommand{\aos}{\jargon{AoS}\xspace}
\newcommand{\saos}{\jargon{SoAoS}\xspace}
\newcommand{\soa}{\jargon{SoA}\xspace}
\newcommand{\randomorder}{\jargon{random-order}\xspace}
\newcommand{\packforce}{\jargon{pack-force}\xspace}
\newcommand{\packgrad}{\jargon{pack-gradient}\xspace}
\newcommand{\packshared}{\jargon{pack-shared}\xspace}
\newcommand{\packforcetype}{\jargon{pack-force-type}\xspace}
\newcommand{\packgradtype}{\jargon{pack-gradient-type}\xspace}
\newcommand{\packsharedtype}{\jargon{pack-shared-type}\xspace}

\newcommand{\none}{\jargon{by-particle}\xspace}
\newcommand{\bystruct}{\jargon{by-struct}\xspace}
\newcommand{\bytype}{\jargon{by-type}\xspace}
\newcommand{\byelement}{\jargon{by-element}\xspace}
\newcommand{\bystructandtype}{\jargon{by-struct-and-type}\xspace}

\newcommand{\FLOAT}{\texttt{float}\xspace}
\newcommand{\DOUBLE}{\texttt{double}\xspace}

\newtheorem{recommendation}{Recommendation}

\begin{document}
\title{Memory Layouts for GPU-Data Transfer Buffering in SPH}
\titlerunning{Memory Layouts for GPU-Data Transfer Buffering in SPH}
%
\author{Mladen Ivkovic\inst{1}\orcidID{0000-0002-3539-3831} \and
Abouzied M.A. Nasar\inst{2}\orcidID{0000-0002-5116-9696} \and
Tobias Weinzierl\inst{1}\orcidID{0000-0002-6208-1841} \and
Matthieu Schaller\inst{3,4}\orcidID{0000-0002-2395-4902} \and
Benedict D. Rogers\inst{2}\orcidID{0000-0002-3269-7979} \and
Georgios Fourtakas\inst{2}\orcidID{0000-0001-8584-3020} \and
Scott T. Kay\inst{5}\orcidID{0000-0002-2277-9049}
}
\authorrunning{M. Ivkovic et al.}
%
\institute{Department of Computer Science, Durham University, South Road, Durham DH1 3LE, UK \and
School of Engineering, The University of Manchester, Nancy Rothwell Building, Manchester M13 9QS, UK \and
Lorentz Institute for Theoretical Physics, Leiden University, PO Box 9506, NL-2300 RA Leiden, the Netherlands \and
Leiden observatory, Leiden University, PO Box 9513, NL-2300 RA Leiden, the Netherlands \and
Jodrell Bank Centre for Astrophysics, Department of Physics and Astronomy, The University of Manchester, Manchester M13 9PL, UK}
%
\maketitle              
\begin{abstract}
The rise in GPU compute speed has outpaced improvements in host-to-device memory transfer speeds, despite the advent of shared-memory superchips. Consequently, memory transfer times now constitute an increasingly large fraction of total time-to-solution, compelling developers to compress GPU kernel input and output data into compact, minimal formats prior to GPU-offloading.
Complementary to existing work on GPU- and compute-friendly data arrangements, this work argues that significant attention should furthermore be paid to the data organisation on the host such that quick data rearrangements and offloading become feasible.
We study a Smoothed Particle Hydrodynamics solver and propose memory layout strategies for host-side particle data that are particularly well-suited to GPU-offloading. Specifically, we advocate splitting classic array-of-struct data structures into a struct-of-arrays-of-structs arrangement, in which each logical struct decomposes into substructs determined by kernel read/write access patterns and attribute types. Splitting a monolithic particle struct into several bespoke, finer-grained structs can reduce the time required to pack data to and from buffers by $\approx 20\%-40\%$, lowering total time spent on GPU-offloading by $\approx 12\%$–$25\%$.

\end{abstract}

\section{Introduction}
\label{section:introduction}

%
%
In 2026, supercomputers rely heavily on general-purpose Graphics Processing Units (GPUs)~\cite{jones2026gaming}.
The accelerators favour specific types of calculations:
they excel when algorithms exhibit regular, contiguous data access and uniform, independent computations.
Most importantly, they favour codes that do not stress the host-GPU interconnect, since it remains slow compared to the GPU on-board memory access, which is typically hitting high-bandwidth memory (HBM).
This does not qualitatively change with the advent of superchips with tightly integrated, shared memory.
Consequently, certain application types are more popular on these systems than others, and data organisation and access patterns become paramount considerations for efficient GPU utilisation.

%
%
%
One algorithmic class not straightforward to port to GPUs are Lagrangian particle methods.
Smoothed Particle Hydrodynamics (SPH) is a prominent example of such a Lagrangian method that is widely used for hydrodynamics simulations in astrophysics since the 1970s~\cite{gingold1977smoothed,lucy1977numerical}.
It's valued for its inherent ability to capture flows with strongly localised features that move in space and time.
In the present paper, we focus on the \swift~astrophysics and cosmology simulation package~\cite{schallerSWIFTModernHighly2024}, which uses SPH to create massive-scale simulations of the Universe \cite{schaye2026colibre,schayeEAGLEProjectSimulating2015,schaye2023flamingo} using hundreds of billions of particles.
Originally developed as a CPU-only solver, porting \swift~to GPUs surfaces many challenges symptomatic of Lagrangian methods in general.

Its core compute kernels describe individual particle updates or particle--particle interactions that can be written in an embarrassingly parallel manner, i.e.~fit to a GPU.
The natural data model for a particle-based code is the Array-of-Structs (AoS).
AoS facilitates particle reordering, sorting, and exchange in a distributed memory environment.
Moreover, it maps naturally onto how domain scientists conceptualise data.
However, AoS is not the preferred layout from a performance standpoint, as fast kernels tend to favour Structs-of-Arrays (SoA).
Beyond that, each SPH compute kernel typically reads and writes only a subset of a particle's attributes.
Looping over particles or particle pairs hence yields scattered data access patterns, while offloading stresses memory bandwidth inappropriately as we ship too much data forth and back~\cite{radtke2026compiler}.
Similar challenges arise in other SPH codes (e.g.~\cite{david2025shamrock,dominguez2022dualsphysics,frontiere2025cosmological}).

%
%
Performance-aware developers therefore optimise their data layouts, most commonly by switching to a Struct-of-Arrays (SoA) representation (e.g. \cite{abelApplicationsTuningStreaming1999}).
A complete rewrite into SoA is, however, problematic in SPH:
optimising for one compute kernel's access pattern introduces performance penalties in other program phases.
Rather than switching persistently, some authors therefore advocate for temporary, on-the-fly data rearrangement.
Such transformations can be applied manually~\cite{nasarTaskparallelismSWIFTHeterogeneous2026} or delegated to a compiler~\cite{radtke2026compiler}.
The latter approach, as well as abstraction layers such as Kokkos (cf.~work around~\cite{thompson2022lammps}), is appealing because data storage decisions need not propagate into the compute kernel code.
They keep domain-specific calculations free of layout concerns.
The precise combination of data formats and reordering steps depends on the exact algorithmic composition, which varies widely with the physics and numerics implemented.
Consequently, hybrid approaches for data layouts such as Array-of-Structs-of-Arrays and Struct-of-Arrays-of-Structs'' (AoSoA and SoAoS, respectively, e.g.~\cite{siegelCUDAMemoryOptimizations}) are used in for example Lattice-Boltzmann methods \cite{shetDataStructureMovement2013, shetVECTORIZATIONLATTICEBASED2013, caloreOptimizationLatticeBoltzmann2019} and particle solvers \cite{mniszewskiEnablingParticleApplications2021}, reporting performance benefits.

However, the literature examines the impact of these hybrid formats on overall code performance or on specific compute kernels.
In the context of SPH, literature explores data layouts for efficient storage of particles in neighbour lists (e.g.~\cite{bilottaOptimizationFlexibleNeighbors2024, thalerGPUNativeCompressedNeighbor2026}), but, to the best of our knowledge, particle data layout variants optimising data conversions and their impact on performance remain undiscussed.
In contrast, the present paper argues that while switching representations on-the-fly can be beneficial, they are, as sole data optimisation, insufficient, as they predominantly focus on kernel compute performance.
Instead, significant attention should also be paid to the data organisation on the host such that quick data rearrangements and offloading become feasible.

%
%
To this end, we propose splitting particles such that each logical particle is represented by multiple physical structs, each holding only a subset of all the attributes.
As each of these sub-structs' data is stored in a separate contiguous array, this corresponds to an SoAoS data layout.
This SoAoS representation can often be designed so that all particle attributes required by a given kernel are grouped into one physical struct, while attributes irrelevant to that calculation remain untouched in separate structs.
Such a split reduces pressure on the memory interconnect.

We further propose that this split can be guided by data-type considerations:
it is sensible to subdivide structs so that one AoS holds exclusively \jargon{double} attributes, while another stores only \jargon{float}s or \jargon{integer}s.
This way, each struct access is homogeneous from a data type point of view which suits vector units.

%
%
Our results focus exclusively on the host-sided data reorganisation for various formats, yet suggest that this struct-splitting strategy can accelerate the total GPU-offloading time by 10\%--25\%.
However, the work also reveals open questions:
it is neither clear how the splitting should be chosen, as the number of potential SoAoS variants grows exponentially with the number of attributes.
Nor is it obvious whether knowledge of the exact layout should feed back into the source code.
Finally, all optimisations presented here are implemented manually.

%
%
The remainder of the paper is organised as follows:
We start with a brief sketch of our software architecture and an algorithmic blueprint, highlighting the points where data transformation becomes performance-critical.
In Section~\ref{section:method}, we present the implementation method to address the data access burden.
Performance results in Section~\ref{section:results} demonstrate the potential of the approach, before a brief summary and outlook in Section~\ref{section:conclusion} concludes the discussion.

\section{Problem Statement}
\label{section:problem-formulation}
\vspace{-0.2cm}

\subsection{SPH Compute Kernels}

In SPH, local fluid quantities at particle positions are mapped onto the particles' properties as convolutions with a weighting kernel.
In discrete form, these become weighted sums over neighbouring particles~\cite{priceSmoothedParticleHydrodynamics2012} and give rise to the algorithmic core of SPH in form of \emph{interaction loops}, i.e.~looping over all neighbouring particles, per reference particle, to collect their weighted contributions.
The exact number of neighbours depends on factors such as the choice of the weighting kernel, the requested spatial resolution, and local conditions of the particles, but is usually $\mathcal{O}(100)$.
In our case, the interaction radius $H$ can vary per particle to accommodate large differences in particle number density \cite{borrowSWIFTMaintainingWeakscalability2018}.

\newcommand{\colwidthPartData}{0.6cm}

\begin{table}[htb]
    \centering
    \caption{
Some particle data and their types used by \swift's~\sphenix SPH flavour.
The last six columns mark whether that variable is used during a \packing or \unpacking operation (column prefix `p-' and `u-', respectively) for the \density, \gradient, and \force interaction loops (suffix `d', `g', and `f', respectively).
    \label{tab:particle_data_exerpt}
    \vspace{-0.2cm}
    }
\resizebox{\linewidth}{!} {
    \begin{tabular}{| l | l | p{6cm} | c | c | c | c | c | c |}
        \hline
        \textbf{Data type}
        & \small{\textbf{Variable name}}
        & \textbf{Semantics}
        & \textbf{p-d}
        & \textbf{u-d}
        & \textbf{p-g}
        & \textbf{u-g}
        & \textbf{p-f}
        & \textbf{u-f} \\
        \hline
        \DOUBLE [3] & x & Coordinates & x &  & x & & x & \\
        \FLOAT [3] & v & Velocity & x &  & x & & x & \\
        \FLOAT [3] & a & Acceleration &  &  & & & & x \\
        \FLOAT & mass & Particle mass & x &  & x & & x & \\
        \FLOAT & h & Smoothing length $H$ & x &  & x & & x & \\
        \FLOAT & u & Internal energy $u$ &  &  & x & & x & \\
        \FLOAT & u\_dt & $\partial u/ \partial t $ &  &  & & & & x \\
        \FLOAT & rho & Density &  & x & x & & x & \\
        \texttt{int8\_t} & time\_bin & particle time step size & & & & & x & \\
        \vdots & \vdots & \vdots & \vdots & \vdots & \vdots & \vdots & \vdots & \vdots
        \\
        \hline
    \end{tabular}
} 
\end{table}

Per time integration step, at least two such interaction loop types are performed:
First, the \density~loop determines the local $H$ neighbourhood and mass densities per particle.
Forces acting on the particles can then be computed in a subsequent \force~loop.
\sphenix~\cite{borrowSphenixSmoothedParticle2022}, the SPH flavour used in this work, employs a third \gradient~loop between the \density~and the \force~loops, and other modern SPH formulations might inject even further ones to increase the method's accuracy and order of convergence.
While all kernels read the particle positions, each  interaction loop type reads and writes a different overall set of the total number of particle attributes (Table~\ref{tab:particle_data_exerpt}).

The pair-wise interaction kernels team up with linear update kernels which simply traverse a set of particles and update some of their attributes.
While the interaction loops yield computationally expensive kernels with a local complexity of $\mathcal{O}(N)$ over the number of particles $N$ within a neighbourhood,
the linear kernels are cheap and ignored from hereon.

\subsection{Data Structures and Tasks in \swift}

Like many SPH solvers \cite{david2025shamrock,frontiere2025cosmological}, \swift~sorts particles into cells of an octree:
The octree is refined up to a leaf cell size of $2 H_{\mathrm{max}}$,
where each cell hosts a set of particles and $H_{\mathrm{max}}$ is the largest interaction radius of any of its particles.
In the vanilla code version, the particles are encoded as a linear sequence, i.e.~the particles are held as AoS.
The refinement threshold on the cell size ensures any particle's neighbours needed during interaction loops are contained either within the cell itself or one of its adjacent cells. 

This constrains the workload per interaction loop and also ensures a sufficient compute efficiency:
For an arbitrary set of particles, only few will actually contribute towards a force or density, respectively, as only few are within the interaction radius.
As particles are stored closely within a cell, the likelihood is high that most particles within a cell actually contribute towards the kernel outcome, while particles that are far away and therefore will not contribute are not considered right from the start.

Another advantage of the decomposition of particles over an octree is the fact that \swift~can map the deployment of kernels acting on individual cells onto individual tasks.
\swift builds up a large task graph describing the SPH scheme, where each individual task represents one step of the algorithmics over a subset of particles.
This leads to high scalability on multicore systems~\cite{schallerSWIFTModernHighly2024}.

\subsection{GPU-Offloading}

While fine-granular tasks over an octree perform well on CPUs, such tasks induce too little work per kernel to benefit from GPU-offloading.
Increasing the cell sizes to contain more particles is not a viable option, as particles only interact with a limited number of neighbours.
It would increase the workload but lead to unnecessary computations and thread divergence.

Rather than immediately computing a task as soon as it is scheduled, \swift's GPU-offloading approach gathers ready tasks into larger bundles of work.
The bundles of tasks are then deployed to a GPU en bloc (cf.~dynamic task fusion in \cite{liDynamicTaskFusion2022}).
The offloading hence decomposes into the following steps:

During the \packing~process, a task's relevant particle data is pre-processed (e.g.~particle coordinates shifted to facilitate periodic boundary conditions) and moved into a buffer on the host (CPU).
As each type of interaction loop (\density, \gradient, and \force) requires only a subset of the particles' data,
only the required quantities are buffered.

Once a bundle reaches a sufficient size, its buffer contents are transferred to the GPU, where the interactions are computed.
   Subsequently, only the strictly required results are transferred back into host-side buffers.
   We refer to the whole sequence of data transfer, compute plus transfer back as a \launch.

Finally, an \unpacking~process performs all required reductions and transfers the host-side result buffer into the actual particles.



\subsection{Target Hardware}

We perform all experiments below on two architectures: A node comprising two Intel Xeon Gold 6430 CPUs (\dine) and a \grace superchip.
The \grace supports CPU-GPU coupling through fast  NVLINK-C2C memory transfer (900Gb/s) compared to the 64Gb/s of the PCIe 4 connected A30.

\subsection{The ``\packing'' Bottleneck and GPU Kernel Realisation}

\begin{figure}[htbp]
  \includegraphics[width=\textwidth]{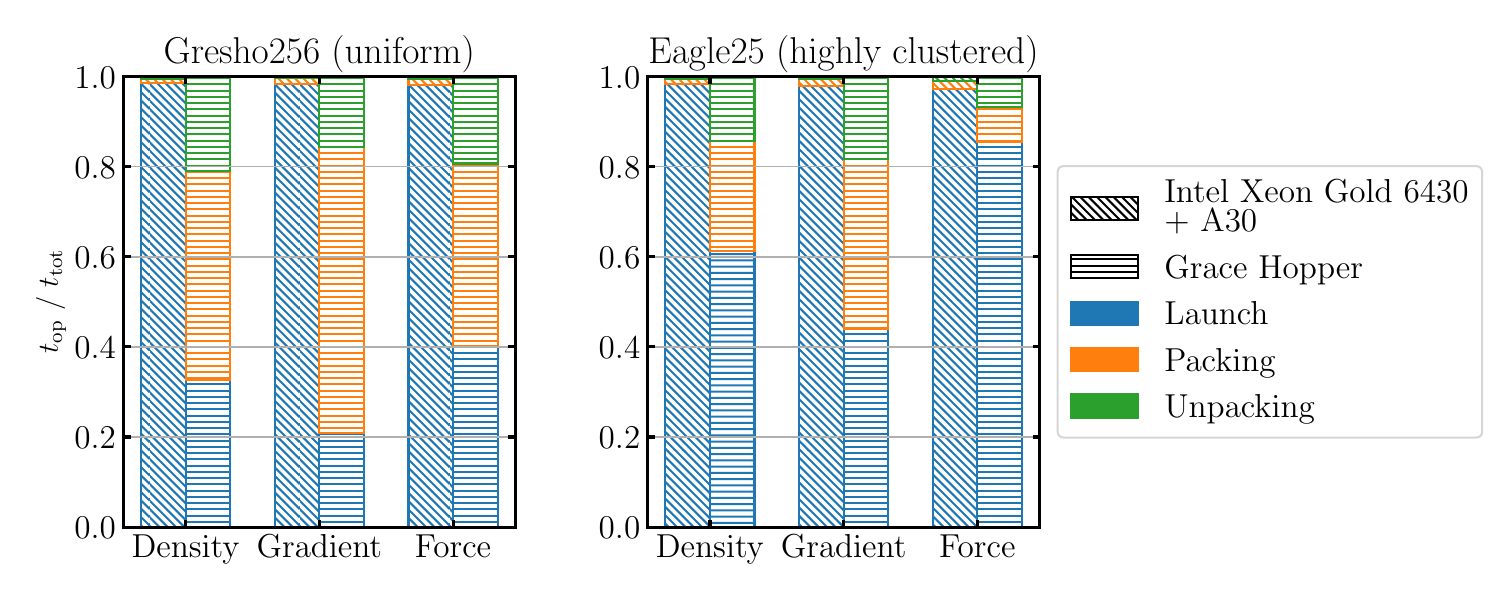}%
  \vspace{-0.6cm}
  \caption{
Relative fraction of time spent in offloading cycles for the interaction loops \density, \gradient, and \force~in each of the three offloading steps (\packing, \launch, \unpacking; blue, orange, and green, respectively.) \swift employs for 16 simulation steps of the \gresho test (left), which has a uniform particle distribution, and \eagle (right), which has a highly clustered particle distribution (see Section~\ref{chap:experiments}).
Timings have been obtained on two architectures: A node comprising two Intel Xeon Gold 6430 CPUs  connected via PCIe 4 to a NVDIA A30 GPU (diagonal hatches) and an NVIDIA Grace Hopper superchip (horizontal hatches).
    \label{fig:pack_timings}
  }%
\end{figure}

Both the \packing and \unpacking steps are executed on the host only, while the \launch process comprises both memory transfers and kernel execution.
Relative fractions of the time spent in offloading cycles for each interaction loop type (\density, \gradient, \force) in each of the three offloading steps (\packing, \launch, \unpacking)
for the \dine and \grace
show how new generations of accelerators diminish the compute time and make the \packing and \unpacking steps the primary bottleneck \cite{nasarTaskparallelismSWIFTHeterogeneous2026}~(Figure~\ref{fig:pack_timings}).
On \dine, the \launch~step accounts for $>97\%$ of the total time, whereas the sum of the fraction of time spent in host-side \packing and \unpacking buffering operations on the \grace rises from $\sim 2\%$ to $\sim 15-55\%$ for the \eagle test (which comprises highly clustered particle distributions commonly encountered in cosmological simulations) and $\sim 60-80\%$ for the \gresho test, (which has a uniform particle distribution and fewer particles per cell on average, see Section~\ref{chap:experiments}).

Additional AoS-to-SoA~conversations are out of scope here, yet would amplify the effect that the GPU kernels start to outpace their memory transfer and packing counterparts.
\swift's current GPU-offloading is realised through native CUDA, but we observe that OpenMP's \texttt{map} clause realises similar \packing and \unpacking.
Switching to logically shared memory in return implies that all data transfer is delegated to the hardware via the equivalent of cache line misses.
While the hardware might be able to overlap memory transfer and computations more efficiently than \swift~with its asynchronous memory streams, it implicitly organises data transfers in memory blocks and hence also moves data which is not required by compute kernels in an AoS~formulation.
In our setups, this makes it slower than explicit memory transfers.

\section{Particle Layout Studies and Optimisation}
\label{section:method}

Given that the GPU compute kernels require AoS data, in the studies presented, \packing means converting AoS data into AoS buffers.
Since only a subset of each particle's data is used during each interaction loop, we hypothesize that an efficient \packing~benefits from removing un-needed struct members within the target buffer, i.e.~the compute kernel benefits from spatial proximity of attributes that are actually used by a kernel and no unncessary data are transferred.

The efficiency of such \packing~hinges upon the data format of the input data.
We therefore experiment with various memory layouts for the host data.

\subsection{Particle Memory Layouts}

\paragraph{Baseline Particle Memory Layouts}

In the \aos version, a single particle's data is held within a single struct.
This mirrors \swift's default \aos~memory arrangement.
The struct members are taken directly from the \sphenix SPH implementation of the \swift~code and are generally arranged in the same order as they are used throughout the code.

To evaluate the influence of the ordering of the members, we also introduce the \randomorder~version, wherein the ordering of the \aos layout's struct members have been randomly shuffled.

\paragraph{Kernel-Driven Particle Memory Layouts}

In an optimised data storage scheme, we split the \aos particle into several structs in anticipation of the kernel access patterns.
The struct members of these Struct-of-Arrays-of-Structs (\saos) variants correspond to the data required for the three interaction types.
However, since a few variables are required in multiple interaction loops, our code offers three subflavours of SoAoS variants.
The \packgrad~variant prioritises the \gradient \packing operation:
All variables required for this operation are placed into its associated struct, disadvantaging the \force \packing loop's access thereof.
Conversely, the \packforce~variant follows the same principle, but prioritises the \force \packing operation instead.
Finally, the \packshared~variant places all variables shared between those two operations in a separate ``shared'' struct.

\paragraph{Data Type-Driven Particle Splitting}

Modern compute units such as AVX vector registers yield their highest throughput if they operate on variables of homogeneous data types for an extended period.
Therefore, we additionally investigate whether they can be improved upon by further splitting each of the \saos structs according to their data types.
These variants are labelled as \packgradtype, \packforcetype, and \packsharedtype~and split their respective baseline further, such that the first substruct hosts only double precision values for example, the second only single precision, then only integers and so forth.

\subsection{Particle Data Accessor Methods}

A simple API hiding the memory layout behind getters and setters allows us to swap particle data memory layout variants at compile time.
For the different \saos~and \soa memory layouts, this pushes the logic of how to access the correct sub-struct into the API implementation (Algorithm~\ref{algorithm:api}),
while the compute kernels can be written storage-agnostic.
The APIs are developed against a pointer to a struct, which internally is broken down into scattered substruct accesses for \saos.
Different API implementations can be generated from templates providing a specification of data layout.

\begin{algorithm}[htb]
\caption{
 The memory layout is realised through plain structs, but all field accesses are hidden behind setters and getters such as \texttt{get\_X}.
 This way, we can swap the memory layout at compile time without altering the computations.
 \label{algorithm:api}
}
\begin{lstlisting}
struct part_data_arrays {      /* holds all global particle data arrays */
  struct part* _part;          /* main particle struct */
  struct subset1* _subset1; /* First subset of particle data */
  struct subset2* _subset2; /* Second subset of particle data */
  /* etc */
};
struct part {                /* main particle struct */
  size_t _index;             /* this particle's index in particle arrays */
  struct part_data_arrays* _global_data; /* access to global p. arrays */
  /* Other variables ... */
};
__attibute__((always_inline)) inline float get_X(const struct part* restrict p) {
  /* Get variable X, where X is stored in a different struct */
  const struct subset1* restrict _subset1 = p->_global_data->_subset1;
  return _subset1[p->_index].X;
}
\end{lstlisting}
\end{algorithm}

Working with logic that maps a (logical) struct pointer onto a meta struct (herein referred to as \partstruct), which then in turn delegates to content therein is appealing as it allows for a wide variety of (potentially hierarchical) memory layouts.
However, it may not be the most efficient access method and it increases the memory footprint per particle (cmp.~additional \jargon{\_index} and \jargon{\_global\_data} attributes). Nevertheless, we include it in our experiments to quantify its associated performance penalties.

We furthermore implement two more versions to be able to assess any access overhead:
The \globalvar accessor variant accesses particle arrays as a global variable.
For this variant, each API call is passed a particle's index within a global arrays instead of the~\jargon{struct part} pointer.
In the second variant (\explicitvar), we explicitly pass a \emph{local} variable which holds pointers to global arrays as an argument to each API call instead of accessing them as a global variable.

\subsection{Loop Fission Variants}

\packing and \unpacking~are computationally challenging as they lack (significant) arithmetics but move and reorder data that is potentially scattered.
We therefore investigate whether these two transformations can be improved by modifying and rearranging their innermost loops.
The baseline \none creates a single loop where the loop body copies all required quantities of a single particle.
The \bystruct variant splits the loop $N$ times, where $N$ is the number of \saos substructs whose data is accessed by the conversion.
Each loop then only accesses a single struct's data.

In the \byelement variant, every single particle quantity being accessed is accessed in a separate loop.
Finally, we also consider data type-driven loop fission.
With the \bytype version, loops are split such that each loop only accesses a single data type.
The \bystructandtype combines this approach with the memory layout driven \bystruct approach and splits the \bystruct loops further by the accessed variable's data types.

\subsection{\swiftgpupacksim~Mini-App}\label{chap:experiments}

To zoom in on the \packing and \unpacking operations, we introduce a purpose-built mini-app ``\swiftgpupacksim''\footnote{\url{https://github.com/mladenivkovic/swiftgpupacksim}}.
It accepts a trace of GPU-offloading actions from a ``real'' \swift~simulation and replays this recorded sequence of \packing~and \unpacking~operations.
Each thread's workload is logged separately over 5 full simulation steps, recording realistic data access patterns of $\mathcal{O}(10^5 - 10^6)$ operations per thread and per simulation step.

The underlying GPU access patterns stem from two different examples.
``\gresho'' consists of a Gresho-Chan vortex \cite{greshoTheorySemiimplicitProjection1990}, where a cubic domain with periodic boundaries is discretised using $256^3 \approx 16.8 \times 10^6$ particles ($\approx 2.5$GiB particle data) which are approximately uniformly spaced in a glass-like distribution.
The minimal, maximal, mean, and median number of particles per cell (and hence per \packing or \unpacking operation) are 55, 73, 64, and 64, respectively.

The second used experiment, ``\eagle'', tests the behaviour and performance on highly non-uniform initial conditions extracted from the EAGLE cosmological simulation campaign \cite{schayeEAGLEProjectSimulating2015} at late cosmic times, where cosmological structures and voids are well-formed, leading to large differences in particle number densities contained within the simulated volume.
It consists of $\approx 50.7 \times 10^6$ particles ($\approx 7.74$GiB of particle data).
Here, the minimal, maximal, mean, and median number of particles per cell are 5, 125'113, 184, and 109, respectively.

\section{Results}
\label{section:results}

\paragraph{Memory Layout Impact on Overall Simulation}

\begin{figure}[htb]
  \includegraphics[width=\textwidth]{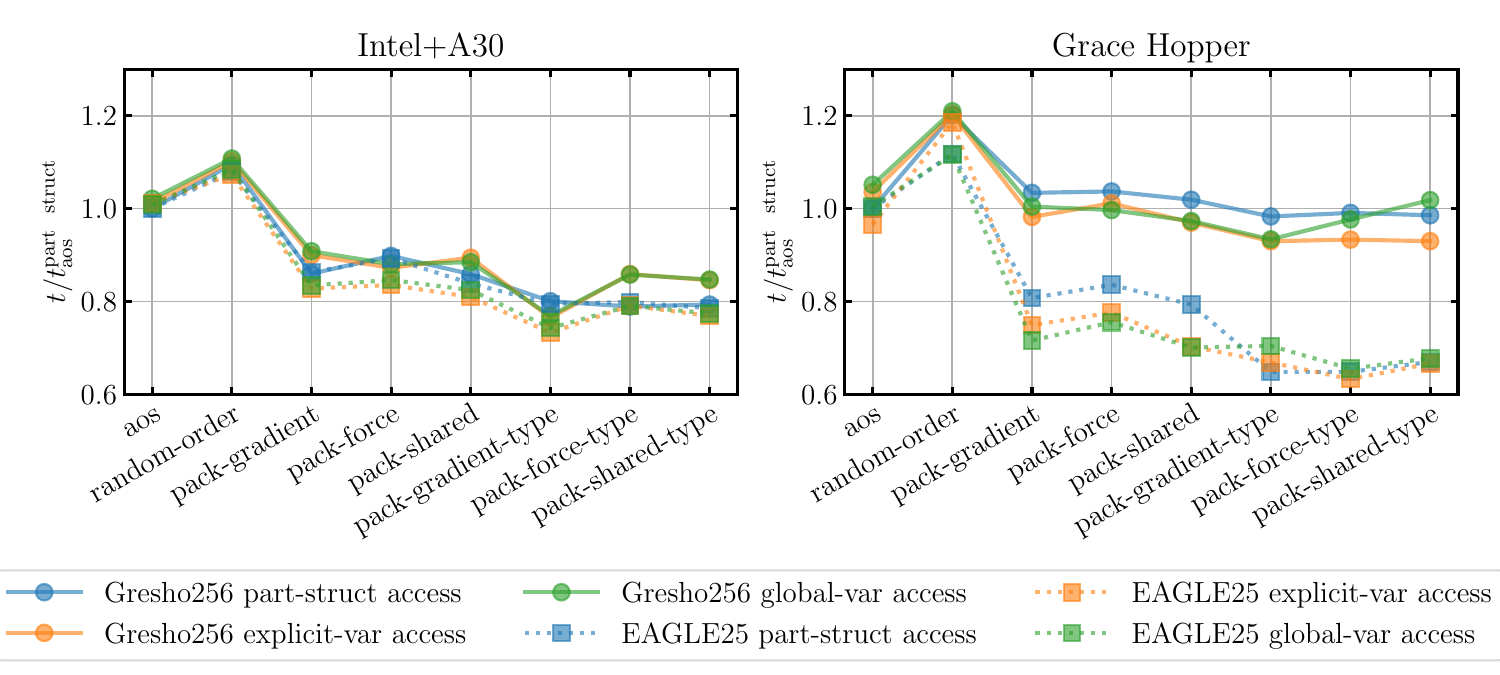}%
  \vspace{-0.4cm}
  \caption{
    Time to complete all \packing and \unpacking operations for varying particle memory layout variants (x-axis) and accessor methods (line colours) over all compute kernel types.
    Results obtained from \dine (left) and  \grace (right) normalised against the \partstruct accessor and \aos baseline.
    \label{fig:part_access_total_times}
  }%
\end{figure}

%
%
We first assess the total time required to complete all \packing and \unpacking operations (Figure~\ref{fig:part_access_total_times}).
A smaller runtime indicates a faster conversion kernel, i.e.~a better outcome.

\noindent
The \randomorder~memory layout consistently demands more time than its well-ordered \aos counterpart, taking $\approx 10\%$ longer on \dine and $\approx 20\%$ on \grace. It is intuitively clear that grouping particle data together is advantageous, as it allows conversion kernels to transfer contiguous attributes en bloc into the GPU data structure---likely within a single cache line.

\vspace{-0.5em}
\begin{recommendation}
The attributes within a struct should be grouped such that those attributes required by one GPU compute kernel sit next to each other.
\end{recommendation}
\vspace{-0.5em}

\noindent
With the exception of the \gresho~experiment on \grace, all \saos variants outperform the \aos variant by up to $\approx 10-20\%$ on \dine and $\approx 20-40\%$ on \grace. Further splitting the \saos variants by data type generally yields additional, modest runtime improvements of $\approx 5\%$.
These gains are most pronounced when the particle-per-cell count grows very high.
\noindent
Hence, beyond grouping particle data together, splitting the data into separate structs (\saos) yields further speedup: even if substructs were not stored contiguously, the probability of unnecessary struct fields being loaded into cache during a conversion remains low.

\vspace{-0.5em}
\begin{recommendation}
Subdividing a struct into substructs is advantageous.
\end{recommendation}
\vspace{-0.5em}

\noindent
Unfortunately, codes such as SPH employ many different compute kernels, so it is not clear from the outset which kernel should dominate the data layout.

\paragraph{Impact on individual compute kernels}

\begin{figure}[h]
\includegraphics[width=\textwidth]{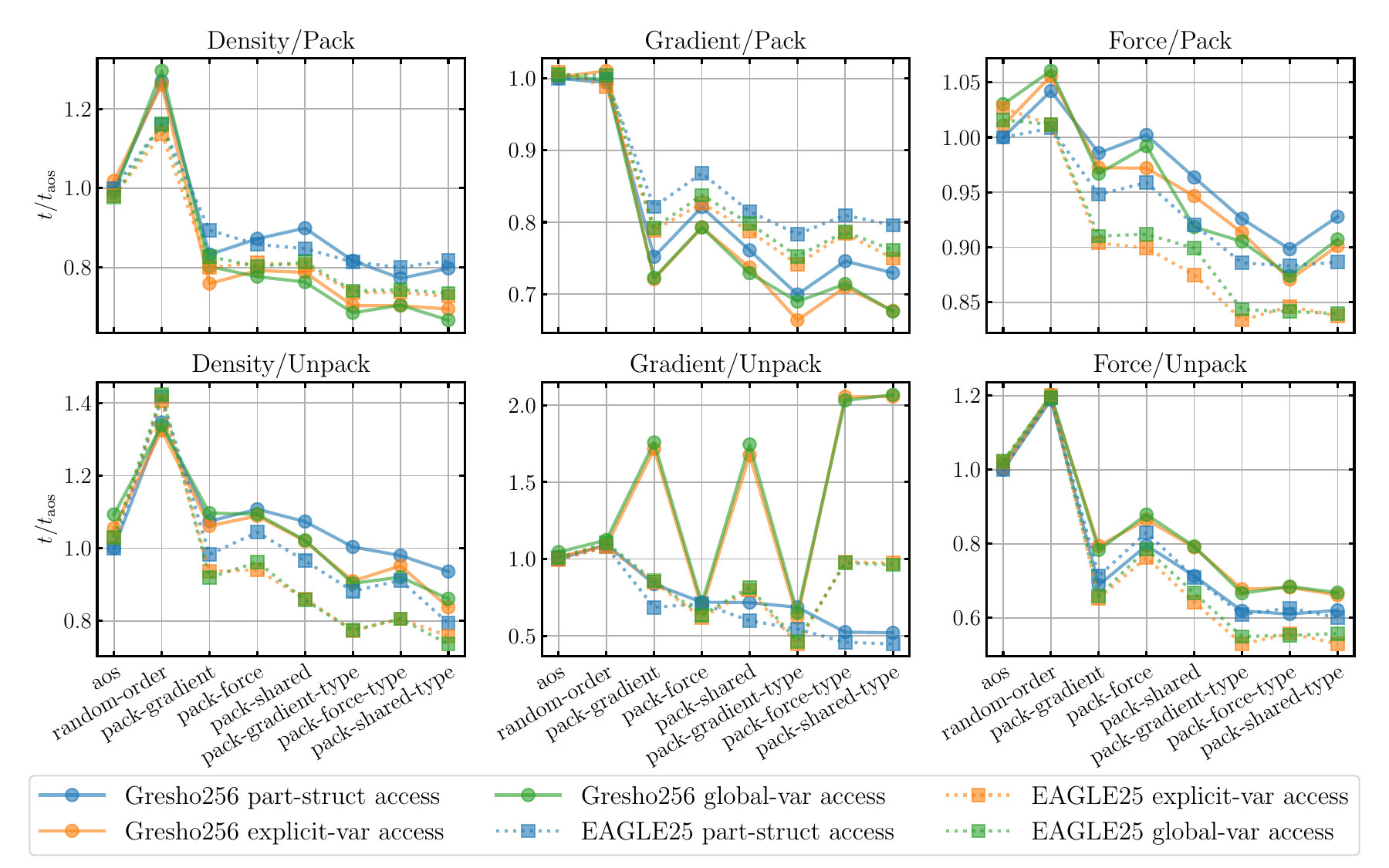}%
\vspace{-0.6cm}
\caption{
  Time to complete all \packing and \unpacking operations for varying particle memory layout variants (x-axis) and accessor methods (line colours) on \dine for different compute kernels  normalised against the \partstruct accessor and \aos baseline.
  \label{fig:part_access}
}%
\end{figure}

We next break down the measurements into the impact of data layouts on individual compute kernels (Figure~\ref{fig:part_access}) and evaluate whether the access realisation affects conversion kernel performance.

\noindent
We observe that certain particle sublayouts have a pronounced effect on different kernels, and no single layout emerges as optimal for all kernels. The total impact of a given layout optimisation is therefore a weighted combination of the runtime improvement due to \saos substructuring and that kernel's contribution to total simulation time.

\vspace{-0.2cm}
\begin{recommendation}
  In an ideal world, the \saos layout becomes kernel dependent, i.e.~changes with the algorithmic phase.
\end{recommendation}
\vspace{-0.2cm}

\noindent
The \explicitvar and \globalvar variants show very little difference from each other, while \partstruct access is consistently the slowest for \packing operations---yet delivers the fastest \gradient and \force \unpacking.
To explain this behaviour, we examined the optimisation reports generated by the compiler (Intel OneApi 2025.2.1).
For the \partstruct accessor, the compiler is unable to auto-vectorise the innermost loop:
the indirection introduced by \saos obscures any contiguous memory access, even when the data is physically stored as a linearised (sub-)AoS stream.
Counterintuitively, this lack of vectorisation sometimes proves beneficial.
The outliers in the \gradient/\unpacking operation (\packgrad, \packshared, \packforcetype, and \packsharedtype) are auto-vectorised, yet vectorisation here hurts performance, likely due to excessive scatters and gathers and a reduced core frequency.

\vspace{-0.5em}
\begin{recommendation}
 As any rationale feeding into the data organisation are not visible to compiler optimisation heuristics, these auto-optimisations can lead to deteriorating performance and have to be studied carefully.
\end{recommendation}
\vspace{-0.5em}

\paragraph{Loop Fission Variants}\label{chap:results_loop_splitting}

\begin{figure}[htb]
\includegraphics[width=\textwidth]{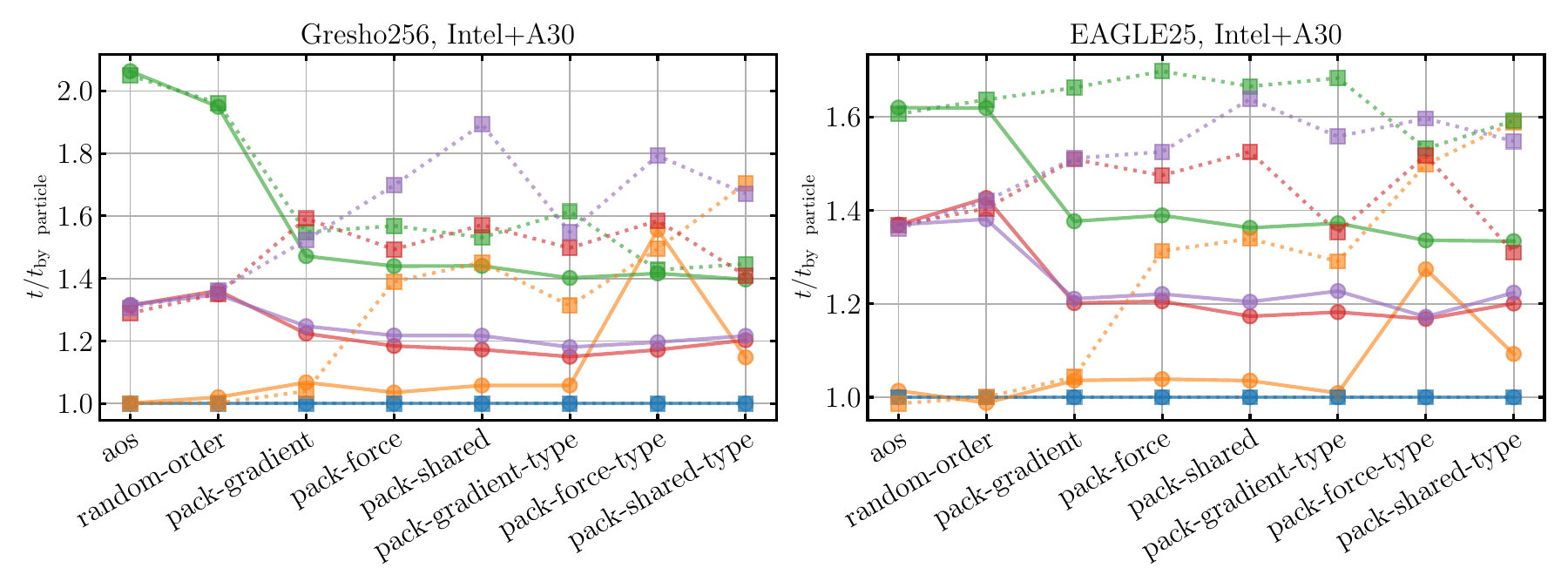}%
\\
\includegraphics[width=\textwidth]{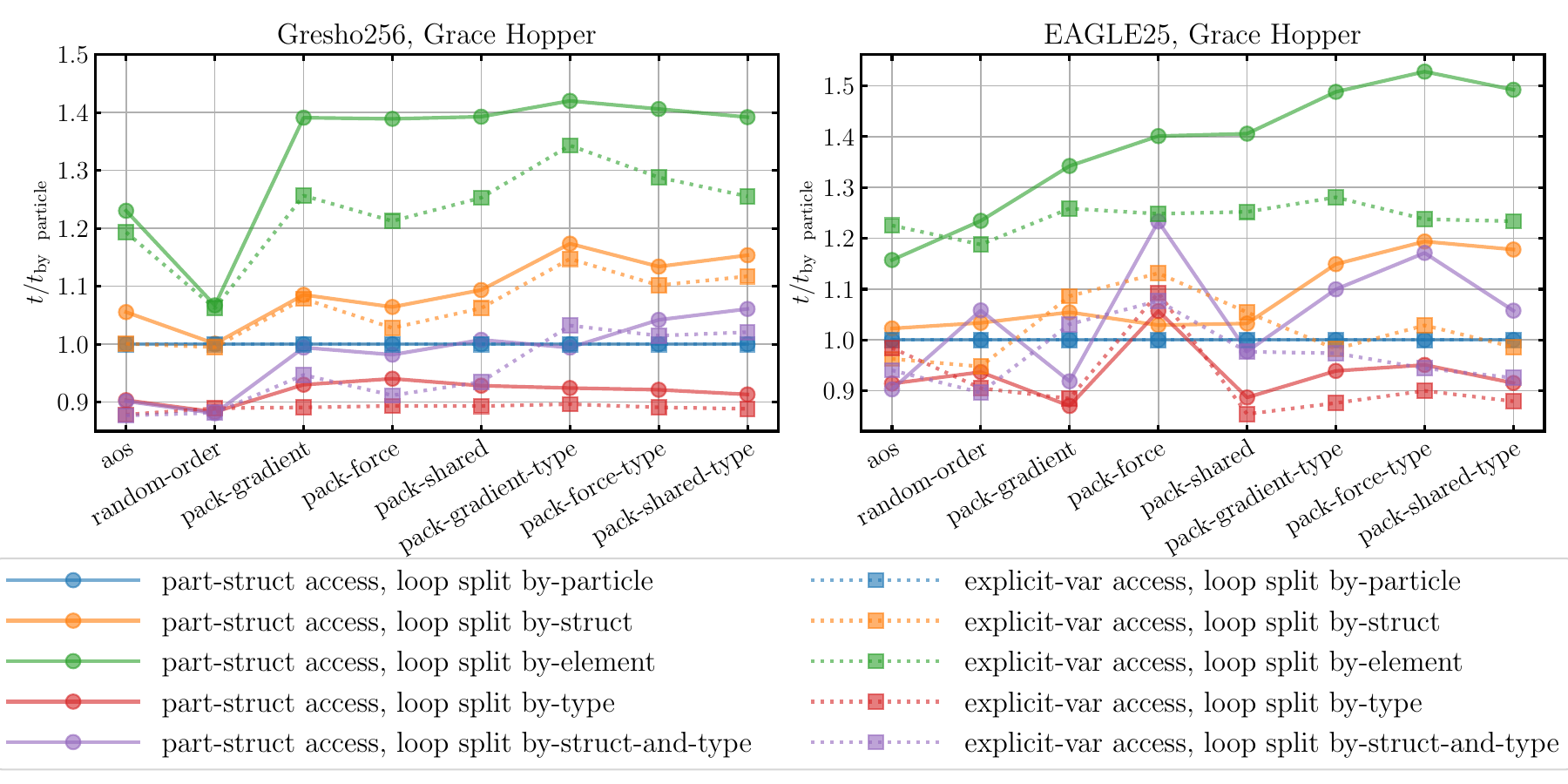}%
  \vspace{-0.2cm}
\caption{
Time required to complete all \packing and \unpacking operations for varying particle memory layout variants (x-axis) and accessor methods (line styles) for different loop fission methods (colours) relative to the \none baseline.
\emph{Top}: results obtained on \dine. \emph{Bottom}: results obtained on \grace.
\emph{Left}: results for the \gresho experiment. \emph{Right}: results for the \eagle experiment.
}%
\label{fig:loop_splitting_relative_times}
\end{figure}

Loop fission yields no improvement on \dine; consulting the optimisation reports, the compiler tends to fuse the loops back together. On \grace, however, splitting loops \bytype reduces runtime by $\approx10\%$ for nearly all memory layouts relative to unsplit loops.
The architecture benefits from homogeneous data-type access patterns (Figure~\ref{fig:loop_splitting_relative_times}).

\vspace{-0.5em}
\begin{recommendation}
  It is advantageous to access different data types in separate loops.
\end{recommendation}
\vspace{-0.5em}

\section{Summary and Outlook}
\label{section:conclusion}

If a code rewrites a data structure prior to invoking a compute kernel on a GPU, managing the conversion overhead is key.
It is natural to convert only those attributes subject to reads and writes by that particular kernel.
In related work, we refer to the subset of affected particle attributes as a \emph{data view}~\cite{radtkeDataViews}.
The present work suggests that identifying a subset of attributes immediately before an on-the-fly conversion is too late.
Instead, we decompose struct data \emph{a priori}, so that a conversion operates on only a subset of attributes from the outset.

Conversely, our data suggest that GPUs' compute capabilities improve at a rate such that the textbook wisdom that SoA plays the dominating role is no longer universally true.
Indeed, many compute kernels on GPUs are now that fast that data transfer dominates the runtime.
This insight challenges traditional AoS-to-SoA conversion techniques~\cite{radtke2026compiler}, but at the same time implies that a proper design of hybrid memory layouts is important.
While the SoAoS technique proves valuable in a robust and consistent manner, fundamental challenges remain.

%
%
First, our implementation of the data structure variants is manual.
This is acceptable for a case study and appropriate for a codebase that solves a single, well-defined problem.
For a general-purpose code such as \swift, however, where users are expected to add further data fields to a struct and to inject new computational kernels (e.g.~to implement additional physics), manually maintaining data conversions is problematic:
It increases code complexity, is error-prone, and complicates the introduction of new features.
Therefore, an open question remains of whether the decomposition of AoS into SoAoS can be hidden behind an abstraction layer such as Kokkos~\cite{carteredwardsKokkosEnablingManycore2014}, injected via compiler directives analogous to attributes or pragmas~\cite{radtke2026compiler}, or fully delegated to the compiler.

%
%
Second, our work does not investigate whether robust heuristics exist or might exist for choosing the SoAoS decomposition.
The minimal reasonable granularity appears to be determined by the view, i.e.~the attributes' read and written by a single compute kernel, though a finer subdivision proves beneficial in some cases if it induces conversion loops operating over homogeneous data types.
Conversely, fusing the attribute sets of multiple kernels into one SoAoS substruct may sometimes be advantageous--for instance, when two kernels inducing different views execute in immediate succession on the same accelerator.
Robust heuristics are needed to guide effective performance engineering in this space.
Data replication may also be worth considering:
if all substructs within an SoAoS decomposition require a particle's position, it may be beneficial to store that information redundantly so that two conversions can proceed independently and without scattered memory access, even though both require the same field.

%
%
Thirdly, our work does not fully resolve the interplay between SoAoS decomposition and vectorisation.
We expect that loop fission and loop reordering can become powerful tools in combination with SoAoS decomposition, particularly in cases where such restructuring allows the compiler to auto-vectorise more aggressively and to overlap the conversion of partial particle data with compute kernel execution.
In those scenarios, a physical decomposition of a logical struct may yield significant speedups.

Finally, our work focuses solely on optimising the conversion between data formats in buffering processes related to GPU-offloading, which we have identified as a primary bottleneck in our current SPH solver implementation. The impact of the proposed optimisations on the overall execution time of the solver remains to be investigated and will be subject of future work. While we expect benefits in some aspects of the solver, such as in other computational kernels omitted from this work and reduced MPI message sizes, other phases, such as binning particles into cells of the octree as they move through the domain, are likely to incur penalties.

\begin{credits}
\subsubsection{\ackname}
The authors would like to thank EPSRC and the UK’s ARCHER2 eCSE GPU grant programme for funding the research in which the solver presented in this paper was developed with grants EP/W026775/1 and ARCHER2-eCSE02-28.

This work used the DiRAC@Durham facility managed by the Institute for Computational Cosmology on behalf of the STFC DiRAC HPC Facility (www.dirac.ac.uk). The equipment was funded by BEIS capital funding via STFC capital grants ST/K00042X/1, ST/P002293/1, ST/R002371/1 and ST/S002502/1, Durham University and STFC operations grant ST/R000832/1. DiRAC is part of the National e-Infrastructure.

\subsubsection{\discintname}
The authors have no competing interests to declare that are relevant to the content of this article.
Large language models have been used to proofread the text, but no text or code has been generated by artificial intelligence.
\end{credits}
%
%
%
\bibliographystyle{splncs04}

\bibliography{CP128_references}

@article{abelApplicationsTuningStreaming1999,
  title = {Applications {{Tuning}} for {{Streaming SIMD Extensions}}},
  author = {Abel, James and Balasubramanian, Kumar and Bargeron, Mike and Craver, Tom and Phlipot, Mike},
  year = 1999,
  journal = {Intel Technology Journal},
  publisher = {Intel Technology Journal},
  langid = {english}
}

@article{bilottaOptimizationFlexibleNeighbors2024,
  title = {Optimization of Flexible Neighbors Lists in {{Smoothed Particle Hydrodynamics}} on {{GPU}}},
  author = {Bilotta, Giuseppe and Zago, Vito and H{\'e}rault, Alexis and Cappello, Annalisa and Ganci, Gaetana and Van Ettinger, Hendrik D. and Dalrymple, Robert A.},
  year = 2024,
  month = oct,
  journal = {Advances in Engineering Software},
  volume = {196},
  pages = {103711},
  issn = {09659978},
  doi = {10.1016/j.advengsoft.2024.103711},
  urldate = {2024-08-11},
  langid = {english}
}

@article{borrowSphenixSmoothedParticle2022,
  title = {{\textsc{Sphenix}} : Smoothed Particle Hydrodynamics for the next Generation of Galaxy Formation Simulations},
  author = {Borrow, Josh and Schaller, Matthieu and Bower, Richard G and Schaye, Joop},
  year = 2022,
  month = feb,
  journal = {Monthly Notices of the Royal Astronomical Society},
  volume = {511},
  number = {2},
  pages = {2367--2389},
  issn = {0035-8711, 1365-2966},
  doi = {10.1093/mnras/stab3166},
  urldate = {2025-09-24},
  copyright = {https://creativecommons.org/licenses/by/4.0/},
  langid = {english}
}

@inproceedings{borrowSWIFTMaintainingWeakscalability2018,
  title = {{{SWIFT}}: {{Maintaining}} Weak-Scalability with a Dynamic Range of 10\textasciicircum 4 in Time-Step Size to Harness Extreme Adaptivity},
  shorttitle = {{{SWIFT}}},
  booktitle = {13th {{SPHERIC International Workshop}}},
  author = {Borrow, Josh and Bower, Richard G. and Draper, Peter W. and Gonnet, Pedro and Schaller, Matthieu},
  year = 2018,
  month = jul,
  eprint = {1807.01341},
  pages = {44--51},
  address = {Galway, Ireland},
  doi = {10.48550/arXiv.1807.01341},
  urldate = {2018-10-03},
  archiveprefix = {arXiv},
  langid = {english}
}

@article{caloreOptimizationLatticeBoltzmann2019,
  title = {Optimization of Lattice {{Boltzmann}} Simulations on Heterogeneous Computers},
  author = {Calore, E and Gabbana, A and Schifano, Sf and Tripiccione, R},
  year = 2019,
  month = jan,
  journal = {The International Journal of High Performance Computing Applications},
  volume = {33},
  number = {1},
  pages = {124--139},
  issn = {1094-3420, 1741-2846},
  doi = {10.1177/1094342017703771},
  urldate = {2026-10-14},
  langid = {english}
}

@article{carteredwardsKokkosEnablingManycore2014,
  title = {Kokkos: {{Enabling}} Manycore Performance Portability through Polymorphic Memory Access Patterns},
  shorttitle = {Kokkos},
  author = {Carter Edwards, H. and Trott, Christian R. and Sunderland, Daniel},
  year = 2014,
  month = dec,
  journal = {Journal of Parallel and Distributed Computing},
  volume = {74},
  number = {12},
  pages = {3202--3216},
  issn = {07437315},
  doi = {10.1016/j.jpdc.2014.07.003},
  urldate = {2024-08-05},
  langid = {english}
}

@article{greshoTheorySemiimplicitProjection1990,
  title = {On the Theory of Semi-Implicit Projection Methods for Viscous Incompressible Flow and Its Implementation via a Finite Element Method That Also Introduces a Nearly Consistent Mass Matrix. {{Part}} 2: {{Implementation}}},
  shorttitle = {On the Theory of Semi-Implicit Projection Methods for Viscous Incompressible Flow and Its Implementation via a Finite Element Method That Also Introduces a Nearly Consistent Mass Matrix. {{Part}} 2},
  author = {Gresho, Philip M. and Chan, Stevens T.},
  year = 1990,
  journal = {International Journal for Numerical Methods in Fluids},
  volume = {11},
  number = {5},
  pages = {621--659},
  issn = {1097-0363},
  doi = {10.1002/fld.1650110510},
  urldate = {2025-09-24},
  langid = {english}
}

@inproceedings{liDynamicTaskFusion2022,
  title = {Dynamic Task Fusion for a Block-Structured Finite Volume Solver over a Dynamically Adaptive Mesh with Local Time Stepping},
  booktitle = {High Performance Computing},
  author = {Li, Baojiu and Schulz, Holger and Weinzierl, Tobias and Zhang, Han},
  editor = {Varbanescu, Ana-Lucia and Bhatele, Abhinav and Luszczek, Piotr and Marc, Baboulin},
  year = 2022,
  pages = {153--173},
  publisher = {Springer International Publishing},
  address = {Cham},
  isbn = {978-3-031-07312-0}
}

@article{mniszewskiEnablingParticleApplications2021,
  title = {Enabling Particle Applications for Exascale Computing Platforms},
  author = {Mniszewski, Susan M and Belak, James and Fattebert, Jean-Luc and Negre, Christian Fa and Slattery, Stuart R and Adedoyin, Adetokunbo A and Bird, Robert F and Chang, Choongseok and Chen, Guangye and Ethier, St{\'e}phane and Fogerty, Shane and Habib, Salman and Junghans, Christoph and {Lebrun-Grandi{\'e}}, Damien and {Mohd-Yusof}, Jamaludin and Moore, Stan G and {Osei-Kuffuor}, Daniel and Plimpton, Steven J and Pope, Adrian and Reeve, Samuel Temple and Ricketson, Lee and Scheinberg, Aaron and Sharma, Amil Y and Wall, Michael E},
  year = 2021,
  month = nov,
  journal = {The International Journal of High Performance Computing Applications},
  volume = {35},
  number = {6},
  pages = {572--597},
  issn = {1094-3420, 1741-2846},
  doi = {10.1177/10943420211022829},
  urldate = {2026-10-14},
  langid = {english}
}

@article{nasarTaskparallelismSWIFTHeterogeneous2026,
  title = {Task-Parallelism in {{SWIFT}} for Heterogeneous Compute Architectures},
  author = {Nasar, Abouzied M A and Rogers, Benedict D and Fourtakas, Georgios and Ivkovic, Mladen and Weinzierl, Tobias and Kay, Scott T and Schaller, Matthieu},
  year = 2026,
  month = jan,
  journal = {RAS Techniques and Instruments},
  volume = {5},
  pages = {rzag008},
  issn = {2752-8200},
  doi = {10.1093/rasti/rzag008},
  urldate = {2026-05-10}
}

@article{priceSmoothedParticleHydrodynamics2012,
  title = {Smoothed {{Particle Hydrodynamics}} and {{Magnetohydrodynamics}}},
  author = {Price, Daniel J.},
  year = 2012,
  month = feb,
  journal = {Journal of Computational Physics},
  volume = {231},
  number = {3},
  eprint = {1012.1885},
  pages = {759--794},
  issn = {00219991},
  doi = {10.1016/j.jcp.2010.12.011},
  urldate = {2020-06-25},
  archiveprefix = {arXiv}
}

@article{schallerSWIFTModernHighly2024,
  title = {{{SWIFT}}: A Modern Highly Parallel Gravity and Smoothed Particle Hydrodynamics Solver for Astrophysical and Cosmological Applications},
  author = {Schaller, Matthieu and Borrow, Josh and Draper, Peter W and Ivkovic, Mladen and McAlpine, Stuart and Vandenbroucke, Bert and Bah{\'e}, Yannick and Chaikin, Evgenii and Chalk, Aidan B G and Chan, Tsang Keung and Correa, Camila and {van~Daalen}, Marcel and Elbers, Willem and Gonnet, Pedro and Hausammann, Lo{\"i}c and Helly, John and Hu{\v s}ko, Filip and Kegerreis, Jacob A and Nobels, Folkert S J and Ploeckinger, Sylvia and Revaz, Yves and Roper, William J and {Ruiz-Bonilla}, Sergio and Sandnes, Thomas D and Uyttenhove, Yolan and Willis, James S and Xiang, Zhen},
  year = 2024,
  month = apr,
  journal = {Monthly Notices of the Royal Astronomical Society},
  volume = {530},
  number = {2},
  pages = {2378--2419},
  issn = {0035-8711, 1365-2966},
  doi = {10.1093/mnras/stae922},
  urldate = {2024-06-20},
  copyright = {https://creativecommons.org/licenses/by/4.0/},
  langid = {english}
}

@article{schayeEAGLEProjectSimulating2015,
  title = {The {{EAGLE}} Project: Simulating the Evolution and Assembly of Galaxies and Their Environments},
  shorttitle = {The {{EAGLE}} Project},
  author = {Schaye, Joop and Crain, Robert A. and Bower, Richard G. and Furlong, Michelle and Schaller, Matthieu and Theuns, Tom and Dalla Vecchia, Claudio and Frenk, Carlos S. and McCarthy, I. G. and Helly, John C. and Jenkins, Adrian and {Rosas-Guevara}, Y. M. and White, Simon D. M. and Baes, Maarten and Booth, C. M. and Camps, Peter and Navarro, Julio F. and Qu, Yan and Rahmati, Alireza and Sawala, Till and Thomas, Peter A. and Trayford, James},
  year = 2015,
  month = jan,
  journal = {Monthly Notices of the Royal Astronomical Society},
  volume = {446},
  pages = {521--554},
  issn = {0035-8711},
  doi = {10.1093/mnras/stu2058},
  urldate = {2018-10-11}
}

@article{shetDataStructureMovement2013,
  title = {Data Structure and Movement for Lattice-Based Simulations},
  author = {Shet, Aniruddha G. and Sorathiya, Shahajhan H. and Krithivasan, Siddharth and Deshpande, Anand M. and Kaul, Bharat and Sherlekar, Sunil D. and Ansumali, Santosh},
  year = 2013,
  month = jul,
  journal = {Physical Review E},
  volume = {88},
  number = {1},
  pages = {013314},
  issn = {1539-3755, 1550-2376},
  doi = {10.1103/PhysRevE.88.013314},
  urldate = {2026-10-14},
  copyright = {http://link.aps.org/licenses/aps-default-license},
  langid = {english}
}

@article{shetVECTORIZATIONLATTICEBASED2013,
  title = {{{ON VECTORIZATION FOR LATTICE BASED SIMULATIONS}}},
  author = {Shet, Aniruddha G. and Siddharth, K. and Sorathiya, Shahajhan H. and Deshpande, Anand M. and Sherlekar, Sunil D. and Kaul, Bharat and Ansumali, Santosh},
  year = 2013,
  month = dec,
  journal = {International Journal of Modern Physics C},
  volume = {24},
  number = {12},
  pages = {1340011},
  issn = {0129-1831, 1793-6586},
  doi = {10.1142/S0129183113400111},
  urldate = {2026-10-14},
  langid = {english}
}

@article{siegelCUDAMemoryOptimizations,
  title = {{{CUDA Memory Optimizations}} for {{Large Data-Structures}} in the {{Gravit Simulator}}},
  author = {Siegel, Jakob and Ributzka, Juergen and Li, Xiaoming},
  year = 2011,
  journal = {Journal of Algorithms \& Computational Technolog},
  volume = {5},
  number = {2},
  doi = {10.1260/1748-3018.5.2.341},
  langid = {english}
}

@misc{thalerGPUNativeCompressedNeighbor2026,
  title = {{{GPU-Native Compressed Neighbor Lists}} with a {{Space-Filling-Curve Data Layout}}},
  author = {Thaler, Felix and Keller, Sebastian},
  year = 2026,
  publisher = {arXiv},
  doi = {10.48550/ARXIV.2602.19873},
  urldate = {2026-10-14},
  copyright = {Creative Commons Attribution Share Alike 4.0 International},
  langid = {english}
}

@article{jones2026gaming,
  title={From Gaming to Science: How the Graphical Processor Unit Became a Supercomputer},
  author={Jones, Matthew L},
  journal={Nuncius},
  volume={1},
  number={aop},
  pages={1--20},
  year={2026},
  publisher={Brill}
}

@article{gingold1977smoothed,
  title={Smoothed particle hydrodynamics: theory and application to non-spherical stars},
  author={Gingold, Robert A and Monaghan, Joseph J},
  journal={Monthly notices of the royal astronomical society},
  volume={181},
  number={3},
  pages={375--389},
  year={1977},
  publisher={Oxford University Press Oxford, UK}
}

@article{lucy1977numerical,
  title={A numerical approach to the testing of the fission hypothesis},
  author={Lucy, Leon B},
  journal={Astronomical Journal, vol. 82, Dec. 1977, p. 1013-1024.},
  volume={82},
  pages={1013--1024},
  year={1977}
}

@article{schaye2023flamingo,
  title={The FLAMINGO project: cosmological hydrodynamical simulations for large-scale structure and galaxy cluster surveys},
  author={Schaye, Joop and Kugel, Roi and Schaller, Matthieu and Helly, John C and Braspenning, Joey and Elbers, Willem and McCarthy, Ian G and Van Daalen, Marcel P and Vandenbroucke, Bert and Frenk, Carlos S and others},
  journal={Monthly Notices of the Royal Astronomical Society},
  volume={526},
  number={4},
  pages={4978--5020},
  year={2023},
  publisher={Oxford University Press}
}

@article{schaye2026colibre,
  title={The COLIBRE project: cosmological hydrodynamical simulations of galaxy formation and evolution},
  author={Schaye, Joop and Chaikin, Evgenii and Schaller, Matthieu and Ploeckinger, Sylvia and Hu{\v{s}}ko, Filip and McGibbon, Robert J and Trayford, James W and Ben{\'\i}tez-Llambay, Alejandro and Correa, Camila and Frenk, Carlos S and others},
  journal={Monthly Notices of the Royal Astronomical Society},
  volume={548},
  number={1},
  pages={stag375},
  year={2026},
  publisher={Oxford University Press}
}

@inproceedings{radtke2026compiler,
  title={Compiler-supported reduced precision and AoS-SoA transformations for heterogeneous hardware},
  author={Radtke, Pawel K and Weinzierl, Tobias},
  booktitle={Proceedings of the 2026 SIAM Conference on Parallel Processing for Scientific Computing (PP)},
  pages={103--116},
  year={2026},
  organization={SIAM}
}

@article{dominguez2022dualsphysics,
  title={DualSPHysics: from fluid dynamics to multiphysics problems},
  author={Dom{\'\i}nguez, Jose M and Fourtakas, Georgios and Altomare, Corrado and Canelas, Ricardo B and Tafuni, Angelo and Garc{\'\i}a-Feal, Orlando and Mart{\'\i}nez-Est{\'e}vez, Ivan and Mokos, Athanasios and Vacondio, Renato and Crespo, Alejandro JC and others},
  journal={Computational Particle Mechanics},
  volume={9},
  number={5},
  pages={867--895},
  year={2022},
  publisher={Springer}
}

@inproceedings{frontiere2025cosmological,
  title={Cosmological Hydrodynamics at Exascale: A Trillion-Particle Leap in Capability},
  author={Frontiere, Nicholas and Emberson, Jeffrey D and Buehlmann, Michael and Rangel, Esteban M and Habib, Salman and Heitmann, Katrin and Larsen, Patricia and Morozov, Vitali and Pope, Adrian and Faucher-Gigu{\`e}re, Claude-Andr{\'e} and others},
  booktitle={Proceedings of the International Conference for High Performance Computing, Networking, Storage and Analysis},
  pages={25--35},
  year={2025}
}

@article{david2025shamrock,
  title={The shamrock code: I--smoothed particle hydrodynamics on GPUs},
  author={David-Cl{\'e}ris, T and Laibe, G and Lapeyre, Y},
  journal={Monthly Notices of the Royal Astronomical Society},
  volume={539},
  number={1},
  pages={1--33},
  year={2025},
  publisher={Oxford University Press}
}

@article{thompson2022lammps,
  title={LAMMPS-a flexible simulation tool for particle-based materials modeling at the atomic, meso, and continuum scales},
  author={Thompson, Aidan P and Aktulga, H Metin and Berger, Richard and Bolintineanu, Dan S and Brown, W Michael and Crozier, Paul S and In't Veld, Pieter J and Kohlmeyer, Axel and Moore, Stan G and Nguyen, Trung Dac and others},
  journal={Computer physics communications},
  volume={271},
  pages={108171},
  year={2022},
  publisher={Elsevier}
}

@ARTICLE{radtkeDataViews,
       author = {{Radtke}, Pawel K. and {Weinzierl}, Tobias},
        title = "{Annotation-guided AoS-to-SoA conversions and GPU offloading with data views in C++}",
      journal = {arXiv e-prints},
         year = 2025,
        month = feb,
          eid = {arXiv:2502.16517},
        pages = {arXiv:2502.16517},
          doi = {10.48550/arXiv.2502.16517},
archivePrefix = {arXiv},
       eprint = {2502.16517},
 primaryClass = {cs.MS},
       adsurl = {https://ui.adsabs.harvard.edu/abs/2025arXiv250216517R}
}
%
%
%
%
%
\end{document}